\documentclass[floatfix,aps,prb,preprint,superscriptaddress]{revtex4}
\usepackage{graphicx}
\usepackage[]{epsfig}
\usepackage{times,amsmath,amssymb}
\begin{document}

\title{Single-electron Spin Resonance in a Quadruple Quantum Dot}

\author{Tomohiro Otsuka}
\email[]{tomohiro.otsuka@riken.jp}
\affiliation{Center for Emergent Matter Science, RIKEN, 2-1 Hirosawa, Wako, Saitama 351-0198, Japan}
\affiliation{Department of Applied Physics, University of Tokyo, Bunkyo, Tokyo 113-8656, Japan}

\author{Takashi Nakajima}
\affiliation{Center for Emergent Matter Science, RIKEN, 2-1 Hirosawa, Wako, Saitama 351-0198, Japan}
\affiliation{Department of Applied Physics, University of Tokyo, Bunkyo, Tokyo 113-8656, Japan}

\author{Matthieu R. Delbecq}
\affiliation{Center for Emergent Matter Science, RIKEN, 2-1 Hirosawa, Wako, Saitama 351-0198, Japan}
\affiliation{Department of Applied Physics, University of Tokyo, Bunkyo, Tokyo 113-8656, Japan}

\author{Shinichi Amaha}
\affiliation{Center for Emergent Matter Science, RIKEN, 2-1 Hirosawa, Wako, Saitama 351-0198, Japan}

\author{Jun Yoneda}
\affiliation{Center for Emergent Matter Science, RIKEN, 2-1 Hirosawa, Wako, Saitama 351-0198, Japan}
\affiliation{Department of Applied Physics, University of Tokyo, Bunkyo, Tokyo 113-8656, Japan}

\author{Kenta Takeda}
\affiliation{Center for Emergent Matter Science, RIKEN, 2-1 Hirosawa, Wako, Saitama 351-0198, Japan}
\affiliation{Department of Applied Physics, University of Tokyo, Bunkyo, Tokyo 113-8656, Japan}

\author{Giles Allison}
\affiliation{Center for Emergent Matter Science, RIKEN, 2-1 Hirosawa, Wako, Saitama 351-0198, Japan}

\author{Takumi Ito}
\affiliation{Center for Emergent Matter Science, RIKEN, 2-1 Hirosawa, Wako, Saitama 351-0198, Japan}
\affiliation{Department of Applied Physics, University of Tokyo, Bunkyo, Tokyo 113-8656, Japan}

\author{Retsu Sugawara}
\affiliation{Center for Emergent Matter Science, RIKEN, 2-1 Hirosawa, Wako, Saitama 351-0198, Japan}
\affiliation{Department of Applied Physics, University of Tokyo, Bunkyo, Tokyo 113-8656, Japan}

\author{Akito Noiri}
\affiliation{Center for Emergent Matter Science, RIKEN, 2-1 Hirosawa, Wako, Saitama 351-0198, Japan}
\affiliation{Department of Applied Physics, University of Tokyo, Bunkyo, Tokyo 113-8656, Japan}

\author{Arne Ludwig}
\affiliation{Angewandte Festk\"orperphysik, Ruhr-Universit\"at Bochum, D-44780 Bochum, Germany}

\author{Andreas D. Wieck}
\affiliation{Angewandte Festk\"orperphysik, Ruhr-Universit\"at Bochum, D-44780 Bochum, Germany}

\author{Seigo Tarucha}%
\affiliation{Center for Emergent Matter Science, RIKEN, 2-1 Hirosawa, Wako, Saitama 351-0198, Japan}
\affiliation{Department of Applied Physics, University of Tokyo, Bunkyo, Tokyo 113-8656, Japan}
\affiliation{Quantum-Phase Electronics Center, University of Tokyo, Bunkyo, Tokyo 113-8656, Japan}
\affiliation{Institute for Nano Quantum Information Electronics, University of Tokyo, 4-6-1 Komaba, Meguro, Tokyo 153-8505, Japan}

\date{\today}
\begin{abstract}
Electron spins in semiconductor quantum dots are good candidates of quantum bits for quantum information processing.
Basic operations of the qubit have been realized in recent years: initialization, manipulation of single spins, two qubit entanglement operations, and readout.
Now it becomes crucial to demonstrate scalability of this architecture by conducting spin operations on a scaled up system.
Here, we demonstrate single-electron spin resonance in a quadruple quantum dot.
A few-electron quadruple quantum dot is formed within a magnetic field gradient created by a micro-magnet.
We oscillate the wave functions of the electrons in the quantum dots by applying microwave voltages and this induces electron spin resonance.
The resonance energies of the four quantum dots are slightly different because of the stray field created by the micro-magnet and therefore frequency-resolved addressable control of the electron spin resonance is possible.
\end{abstract}

\maketitle

\section{INTRODUCTION}

Electron spins in semiconductor quantum dots (QDs) have relatively long coherence times in solid state devices~\cite{2011BluhmNatPhys, 2014ShulmanNatCom, 2014KawakamiNatNano, 2014VeldhorstNatNano} and potential scalability by utilizing the current extensive semiconductor fabrication techniques.
They are considered good candidates for quantum bits~\cite{1998LossPRA} in quantum information processing~\cite{2000NielsenBk, 2010LaddNat}. 
The required elementary operations on the spin-1/2 qubits for quantum information processing have been demonstrated recently.
The spin states are initialized and read out using the Pauli spin blockade (PSB)~\cite{2002OnoSci} or tunneling to the leads from Zeeman split energy levels~\cite{2004ElzermanNature, 2011NowackScience}.
Rotation of single spins has been realized by electron spin resonance (ESR)~\cite{2006KoppensNat}.
Addressability and the speed of single spin rotation are improved by micro-magnet (MM) induced ESR~\cite{2006TokuraPRL, 2008LadriereNatPhys}. 
High-fidelity single-spin rotation decoupled from the fluctuating nuclear spin environment was demonstrated~\cite{2014YonedaPRL}.
Entanglement operations of two spins are realized by utilizing exchange interaction and fast two qubit operations have been demonstrated~\cite{2005PettaSci, 2011BrunnerPRL}.
This scheme of the spin-1/2 qubit is applicable to a wide variety of materials including Si, which has a long spin coherence time~\cite{2014KawakamiNatNano, 2014VeldhorstNatNano}.

Scale up of the QD system is crucial to realize larger scale quantum gate operations and also explore multi-spin physics.
To this end, spin qubit experiments on multiple QDs have been reported in recent years.
In triple QDs, PSB has been observed~\cite{2013KobayashiArx, 2013AmahaPRL} and the exchange only qubit utilizing a triple QD as a single qubit has been demonstrated~\cite{2010LairdPRB, 2012GaudreauNatPhys, 2013MedfordNatNano}.
Towards three spin-1/2 qubits~\cite{2010TakakuraAPL}, ESR in a triple QD was recently realized~\cite{2015Nakajima}.
Experiments on quadruple QDs (QQDs) have also been started~\cite{2012ThalineauAPL, 2014TakakuraAPL}, and a QQD is utilized for realization of two qubit operations on singlet-triplet qubits~\cite{2012ShulmanSci}.
For four spin-1/2 qubits, the precise charge state control in a tunnel coupled QQD has been demonstrated in the few-electron regime~\cite{2014DelbecqAPL}.

In this paper, we demonstrate four distinctly addressable electron spin resonances in a QQD.
First, we realize few-electron charge states in a QQD required to observe PSB.
Second, we observe PSB for readout of ESR signals by utilizing spin rotation induced by the nuclear spins and the MM.
Finally, we observe four ESR signals corresponding to the four individual spins in the QQD.

\section{EXPERIMENTS AND RESULTS}
\subsection{DEVICE AND MEASUREMENT SETUP}

\begin{figure}
\begin{center}
  \includegraphics{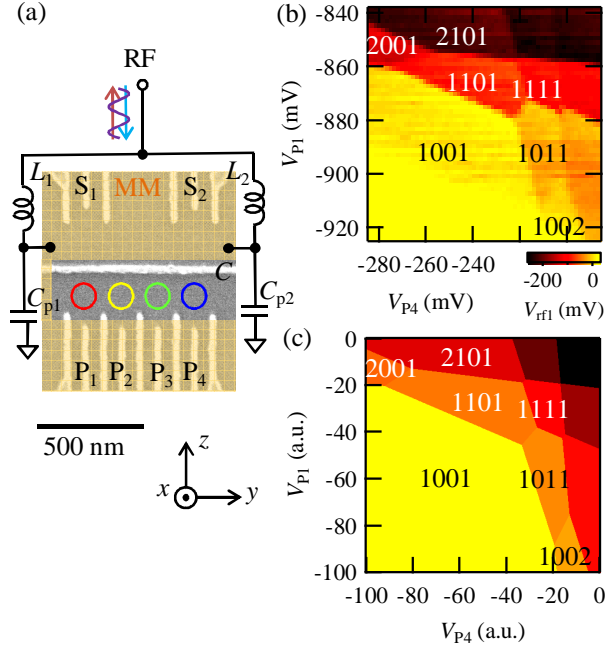}
  \caption{(a) Scanning electron micrograph of the device and the schematic of the measurement setup.
  A QQD is formed at the lower side and the charge states are probed by the charge sensor QDs at the upper side.
  The charge sensors are connected to resonators formed by the inductors $L_{\rm 1}$ and $L_{\rm 2}$ and the stray capacitances $C_{\rm p1}$ and $C_{\rm p2}$ for the RF reflectometry.
  A MM is deposited on the shaded region on the top of the device, which creates local magnetic fields to induce ESR.
  The external magnetic field is applied in plane along the $z$ axis.
  (b) $V_{\rm rf1}$ as a function of $V_{\rm P4}$ and $V_{\rm P1}$.
  Changes of the charge states are observed.
  The number of the electrons in each QD is shown as $n_{\rm 1}, n_{\rm 2}, n_{\rm 3}, n_{\rm 4}$.
  (c) Calculated charge stability diagram of a QQD. 
The experimental result (b) is reproduced by considering the capacitively coupled QQD model.
$n_{\rm 1}, n_{\rm 2}, n_{\rm 3}, n_{\rm 4}$ are shown in the figure.
  }
  \label{Device}
\end{center}
\end{figure}

Figure~\ref{Device}(a) shows a scanning electron micrograph of the device.
The device was fabricated from a GaAs/AlGaAs heterostructure wafer with an electron sheet carrier density of 2.0~$\times$~10$^{15}$~m$^{-2}$ and a mobility of 110~m$^2$/Vs at 4.2~K, measured using the Hall-effect in the van der Pauw geometry.
The two-dimensional electron gas is formed 90~nm under the wafer surface.
We patterned a mesa by wet-etching and formed Ti/Au Schottky surface gates by metal deposition, which appear white in Fig.~\ref{Device}(a).
By applying negative voltages on the gate electrodes, a QQD and two QD charge sensors~\cite{2010BarthelPRB} are formed at the lower and the upper sides, respectively.
The QD charge sensors are connected to RF resonators formed by the inductors $L_{\rm 1}$ and $L_{\rm 2}$ and the stray capacitances $C_{\rm p1}$ and $C_{\rm p2}$ (resonance frequency $f_{\rm res1}$=298~MHz, $f_{\rm res2}$=207~MHz) for the RF reflectometry~\cite{2010BarthelPRB, 1998SchelkopfSci, 2007ReillyAPL}.
The number of electrons in each QD $n_{\rm 1}, n_{\rm 2}, n_{\rm 3},$ and $n_{\rm 4}$ is monitored by the intensity of the reflected RF signal $V_{\rm rf1}$ and $V_{\rm rf2}$.
A MM is deposited on the shaded region on the top of the device, which creates local magnetic fields to induce ESR.
The external magnetic field is applied in the plane along the $z$ axis and magnetizes the MM.
The following measurements were conducted in a dilution fridge cryostat at a temperature of 13~mK.

\subsection{CHARGE STATES}

Figure~\ref{Device}(b) is the charge stability diagram of the QQD.
We measured $V_{\rm rf1}$ as a function of the plunger gate voltages of QD$_{4}$ $V_{\rm P4}$ and QD$_{1}$ $V_{\rm P1}$.
We observe the change of $V_{\rm rf1}$, as the result of the change of the charge states in the QQD.
Charge transition lines with four different slopes are observed reflecting the different electrostatic coupling of the QQD to $V_{\rm P4}$ and $V_{\rm P1}$.
$n_{\rm 1}, n_{\rm 2}, n_{\rm 3},$ and $n_{\rm 4}$ are assigned as shown in Fig.~\ref{Device}(b) by counting the number of charge transition lines from the fully depleted condition [$n_{\rm 1}, n_{\rm 2}, n_{\rm 3}, n_{\rm 4}$]=[0,0,0,0].
Figure~\ref{Device}(c) shows the calculated charge state of the QQD.
By considering the capacitively coupled QQD model, we reproduce the observed charge stability diagram.
We find the characteristic ``goggle" structure, which is formed by the charge transition lines around [1,1,1,1], [1,1,0,1] and [1,0,1,1] charge states.
In the [1,1,1,1] state, each dot contains a single electron and this state is useable as a four qubit system of the spin-1/2 qubit.

\subsection{SPIN BLOCKADE}

\begin{figure}
\begin{center}
  \includegraphics{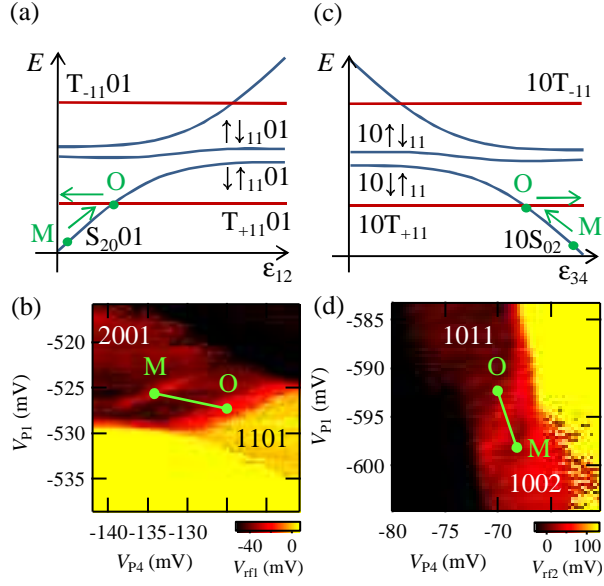}
  \caption{(a), ((c))Energy diagrams and schematics of the pulse operation to observe PSB in QD1 and QD2 (QD3 and QD4).
  The T$_{\rm +11}$01 (10T$_{\rm +11}$) component is formed at the operation point O by using the S$_{\rm 11}01\Leftrightarrow $T$_{\rm +11}01$ (10S$_{\rm 11}\Leftrightarrow $10T$_{\rm +11}$) mixing.
  The triplet component is observed as the [1,1,0,1] ([1,0,1,1]) charge state at the measurement point M.
  (b), ((d)) Observed $V_{\rm rf1}$ ($V_{\rm rf2}$) as a function of $V_{\rm P4}$ and $V_{\rm P1}$.
  The pulse sequences are indicated by lines in the figures.  
  The change of $V_{\rm rf1}$ ($V_{\rm rf2}$) as the result of the spin blocked signals are observed around M.
  }
  \label{SpinBlockade}
\end{center}
\end{figure}

To readout the spin states of the qubits, PSB~\cite{2002OnoSci} is a powerful tool.
The energy of the triplet spin states are higher than that of the singlet in the [2,0,0,1] ([1,0,0,2]) charge region in relatively small magnetic fields.
If the triplet is formed, the charge transition [1,1,0,1]$\rightarrow $[2,0,0,1] ([1,0,1,1]$\rightarrow $[1,0,0,2]) is forbidden.
In the stability diagrams in Figs.~\ref{Device}(b) and (c), the spin blockade can be expected around the charge transition lines between [1,1,0,1] and [2,0,0,1], and between [1,0,1,1] and [1,0,0,2].

We apply voltage pulses on $V_{\rm P1}$ and $V_{\rm P4}$ to observe spin blocked states.
The operation schematics are shown in Figs~\ref{SpinBlockade}(a) and (c).
We apply an external magnetic field of 0.5~T to suppress the effect of the nuclear spins on the PSB~\cite{2005KoppensSci}.
We start from the ground singlet state in QD$_{1}$ S$_{\rm 20}$01 (in QD$_{4}$ 10S$_{\rm 02}$).
The triplet plus component T$_{+11}$01 in QD$_{1}$ and QD$_{2}$ (10T$_{+11}$ in QD$_{3}$ and QD$_{4}$) is formed at the operation point O by using the singlet-triplet mixing S$_{\rm 11}01\Leftrightarrow $T$_{\rm +11}01$ (10S$_{\rm 11}\Leftrightarrow $10T$_{\rm +11}$) induced by the nuclear spins and the MM stray magnetic fields~\cite{2014ChesiPRB}.
At the measurement point M, the triplet components stay in the [1,1,0,1] ([1,0,1,1]) charge state because of PSB and the singlet components relax to the [2,0,0,1] ([1,0,0,2]) charge state.
Then, this blockade can be observed as the change of $V_{\rm rf1}$ ($V_{\rm rf2}$).

Figures~\ref{SpinBlockade}(b) and (d) show the observed $V_{\rm rf1}$ ($V_{\rm rf2}$) as a function of $V_{\rm P4}$ and $V_{\rm P1}$.
We apply voltage pulses with fixed amplitudes as shown as lines in Figs.~\ref{SpinBlockade} (b) and (d).
The directions of the pulses on the stability diagrams are chosen to modulate the detuning, the energy difference of the levels between QD$_{1}$ and QD$_{2}$ (between QD$_{3}$ and QD$_{4}$).
Note that we are also able to control QD$_{2}$ and QD$_{3}$ by $V_{\rm P1}$ and $V_{\rm P4}$ because of the finite capacitive coupling.
Sensor 1 is used for Fig.~\ref{SpinBlockade}(b) (Sensor 2 for Fig.~\ref{SpinBlockade}(d)) to maximize the charge sensitivity.
The changes of $V_{\rm rf1}$ ($V_{\rm rf2}$) are observed around M when the operation point O hits the singlet-triplet mixing point.
These correspond to the spin blocked signals.

\subsection{ELECTRON SPIN RESONANCE}

\begin{figure}
\begin{center}
  \includegraphics{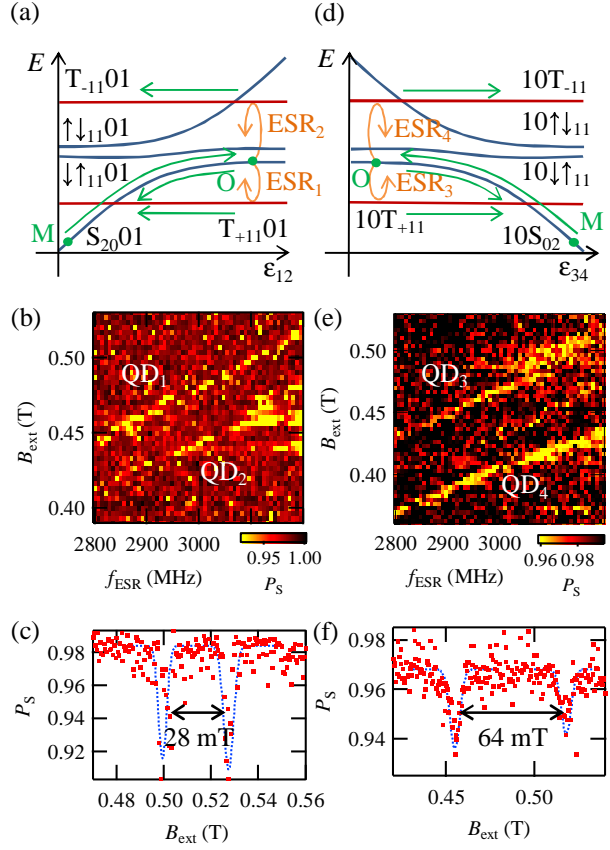}
  \caption{(a), ((d)) Schematics of the energy diagrams and the pulse operations to observe ESR in QD1 and QD2 (QD3 and QD4).
  States are prepared as $\downarrow \uparrow_{11}01$ ($10\downarrow \uparrow_{11}$) due to $\Delta B_{z}$ between QD$_{1}$ and QD$_{2}$ (QD$_{3}$ and QD$_{4}$) by the MM, which is larger than the singlet-triplet splitting at the operation point O.
  The states evolve into T$_{+11}$01 or T$_{-11}01$ (10T$_{+11}$ or 10T$_{-11}$) states by ESR.
The created triplet components are observed as [1,1,0,1] ([1,0,1,1]) charge states at the measurement point M.
  (b), ((e)) Observed $P_{\rm S}$ as a function of $f_{\rm ESR}$ and $B_{\rm ext}$. 
  ESR occurs when the applied microwave frequency matches the external magnetic field plus the stray field created by the micro-magnet $hf_{\rm ESR}=g\mu (B_{\rm ext}+B_{{\rm MM}z})$.
  (c), ((f)) Observed $P_{\rm S}$ as a function of $B_{\rm ext}$ at $f_{\rm ESR}=3265$~MHz.
  Dips of $P_{\rm S}$ are observed when ESR occurs.
  The dips are separated by 28~mT (64~mT) in (c), ((e))
  The dotted curves are Gaussian eye guides.
  }
  \label{ESR}
\end{center}
\end{figure}

Next, we apply a microwave voltage on gate C to induce ESR (with the frequency $f_{\rm ESR}$).
The operation schematics are shown in Figs.~\ref{ESR}(a) and (d).
In the present device, the Zeeman field difference $\Delta B_{z}$ between QD$_{1}$ and QD$_{2}$ (QD$_{3}$ and QD$_{4}$) by the MM will be larger than the singlet-triplet splitting at the operation point O and the eigenstates are $\downarrow \uparrow_{11}$01 and $\uparrow \downarrow_{11}$01 (10$\downarrow \uparrow_{11}$ and 10$\uparrow \downarrow_{11}$), not S$_{11}$01 and T$_{0~11}$01 (10S$_{11}$ and 10T$_{0~11}$).
We prepare the states $\downarrow \uparrow_{11}$01 in QD$_{1}$ and QD$_{2}$ (10$\downarrow \uparrow_{11}$ in QD$_{3}$ and QD$_{4}$) by applying pulses.
Then, we apply microwaves at the operation point O.
These applied microwaves create an oscillating electric field around the gate C and thus induce movements of the wave functions of the QD electrons.
These oscillations of the wave functions are converted into oscillating magnetic fields along the $x$ axis perpendicular to the external magnetic field in the field gradient created by the MM and ESR is induced~\cite{2006TokuraPRL, 2008LadriereNatPhys}.
The triplet components T$_{+11}$01 or T$_{-11}$01 (10T$_{+11}$ or 10T$_{-11}$) are created by ESR and detected as the [1,1,0,1] ([1,0,1,1]) charge states.

Figures~\ref{ESR}(b) and (e) show the singlet return probability $P_{\rm S}$ as a function of $f_{\rm ESR}$ and the external magnetic field $B_{\rm ext}$.
$P_{\rm S}$ is calculated from $V_{\rm rf1}$ ($V_{\rm rf2}$) by using the method reported in the references~\cite{2009BarthelPRL, 2013MedfordNatNano}.
We can see the decrease of $P_{\rm S}$ when the applied microwave frequency matches the external magnetic field plus the $z$ component of the stray field created by the MM $hf_{\rm ESR}=g\mu (B_{\rm ext}+B_{{\rm MM}z})$.
The ESR dips of $P_{\rm S}$ are also observed in Figs.~\ref{ESR} (c) and (f), which show $P_{\rm S}$ as a function of $B_{\rm ext}$ at $f_{\rm ESR}=3265$~MHz.
The dips are separated by 28~mT (64~mT) in Fig.~\ref{ESR}(c) ((f)).

The slopes of the ESR lines in Figs.~\ref{ESR}(b) and (d) give a value of the g-factor as $|g|=0.37$ that is consistent with reported values in previous experiments~\cite{2003PotokPRL, 2003HansonPRL, 2005BeverenNJP}.

We realize addressable control of the operation by choosing appropriate $B_{\rm ext}$ and $f_{\rm ESR}$ such that the separation of the ESR dips is larger than their width as in Figs.~\ref{ESR}(b) and (e).
The intercepts of the ESR lines correspond to the local Zeeman field created by the MM.
From Figs.~\ref{ESR}(b) and (e), the local Zeeman field differences between the quantum dots $B_{{\rm MM}z12}, B_{{\rm MM}z13}, B_{{\rm MM}z14}$ are evaluated as $B_{{\rm MM}z12}=28$~mT, $B_{{\rm MM}z13}=9$~mT, $B_{{\rm MM}z14}=73$~mT.
If there is no misalignment of the QD positions, $B_{{\rm MM}z12}<B_{{\rm MM}z13}<B_{{\rm MM}z14}$ is expected from the design of the MM~\cite{2015YonedaAPE}.
This discrepancy is attributed to the misalignment of the QD positions from the center of the MM.
The observed values of the local Zeeman field are explained by shifts of the QD positions of around 50~nm in the $z$ direction, which is possible in this QQD device.

\section{SUMMARY}

In conclusion, we have demonstrated formation of few-electron charge states, and observed spin blockade and four distinct ESR signals in a QQD.
The four observed ESR dips are well separated and we are able to individually address spins by choosing the appropriate $B_{\rm ext}$ and $f_{\rm ESR}$.
These results will be important for Rabi measurements of four or more spin-1/2 qubits, multiple qubit operations, and demonstration of larger scale quantum gate operations.
These also contribute to exploring multi-spin physics in controlled artificial systems.

\section{ACKNOWLEDGEMENTS}

We thank J. Beil, J. Medford, F. Kuemmeth, C. M. Marcus, D. J. Reilly, K. Ono, RIKEN CEMS Emergent Matter Science Research Support Team and Microwave Research Group in Caltech for fruitful discussions and technical supports.
Part of this work is supported by the Grant-in-Aid for Research Young Scientists B, Funding Program for World-Leading Innovative R\&D on Science and Technology (FIRST) from the Japan Society for the Promotion of Science, ImPACT Program of Council for Science, Technology and
Innovation, Toyota Physical \& Chemical Research Institute Scholars, RIKEN Incentive Research Project, Yazaki Memorial Foundation for Science and Technology Research Grant, Japan Prize Foundation Research Grant, Advanced Technology Institute Research Grant, the Murata Science Foundation Research Grant, and IARPA project ``Multi-Qubit Coherent Operations'' through Copenhagen University. A.L. and A.D.W. acknowledge support of Mercur  Pr-2013-0001, DFG-TRR160,  BMBF - Q.com-H  16KIS0109, and the DFH/UFA  CDFA-05-06.


\begin{references}

\bibitem{2011BluhmNatPhys}
H. Bluhm, S. Foletti, I. Neder, M. Rudner, D. Mahalu, V. Umansky, and A. Yacoby,
{\it Dephasing time of GaAs electron-spin qubits coupled to a nuclear bath exceeding 200$\mu $s},
Nat. Phys. {\bf 7}, 109 (2011).

\bibitem{2014ShulmanNatCom}
M. D. Shulman, S. P. Harvey, J. M. Nichol, S. D. Bartlett, A. C. Doherty, V. Umansky, and A. Yacoby, 
{\it Suppressing qubit dephasing using real-time Hamiltonian estimation},
Nat. Commun. {\bf 5}, 5156 (2014).

\bibitem{2014KawakamiNatNano}
E. Kawakami, P. Scarlino, D. R. Ward, F. R. Braakman, D. E. Savage, M. G. Lagally, M. Friesen, S. N. Coppersmith, M. A. Eriksson, and L. M. K. Vandersypen,
{\it Electrical control of a long-lived spin qubit in a Si/SiGe quantum dot}
Nat. Nano. {\bf 9}, 666 (2014).

\bibitem{2014VeldhorstNatNano}
M. Veldhorst, J. C. C. Hwang, C. H. Yang, A. W. Leenstra, B. de Ronde, J. P. Dehollain, J. T. Muhonen, F. E. Hudson, K. M. Itoh, MorelloA, and A. S. Dzurak,
{\it An addressable quantum dot qubit with fault-tolerant control-fidelity},
Nat. Nano. {\bf 9}, 981 (2014).

\bibitem{1998LossPRA}
D. Loss, and D. P. DiVincenzo,
{\it Quantum computation with quantum dots},
Phys. Rev. A {\bf 57}, 120 (1998).

\bibitem{2000NielsenBk}
M. A. Nielsen, and I. L. Chuang,
{\it Quantum Computation and Quantum Information},
(Cambridge University Press, 2000).

\bibitem{2010LaddNat}
T. D. Ladd, F. Jelezko, R. Laflamme, Y. Nakamura, C. Monroe, and J. L. O'Brien,
{\it Quantum computers},
Nature {\bf 464}, 45 (2010).

\bibitem{2002OnoSci}
K. Ono, D. G. Austing, Y. Tokura, and S. Tarucha,
{\it Current Rectification by Pauli Exclusion in a Weakly Coupled Double Quantum Dot System},
Science {\bf 297}, 1313 (2002).

\bibitem{2004ElzermanNature}
J. M. Elzerman, R. Hanson, L. H. Willems van Beveren, B. Witkamp, L. M. K. Vandersypen, and L. P. Kouwenhoven,
{\it Single-shot read-out of an individual electron spin in a quantum dot},
Nature {\bf 430}, 431 (2004).

\bibitem{2011NowackScience}
K. C. Nowack, M. Shafiei, M. Laforest, G. E. D. K. Prawiroatmodjo, L. R. Schreiber, C. Reichl, W. Wegscheider, and L. M. K. Vandersypen,
{\it Single-Shot Correlations and Two-Qubit Gate of Solid-State Spins},
Science {\bf 333}, 1269 (2011).

\bibitem{2006KoppensNat}
F. H. L. Koppens, C. Buizert, K. J. Tielrooij, I. T. Vink, K. C. Nowack, T. Meunier, L. P. Kouwenhoven, and L. M. K. Vandersypen,
{\it Driven coherent oscillations of a single electron spin in a quantum dot},
Nature {\bf 442}, 766-771 (2006).

\bibitem{2006TokuraPRL}
Y. Tokura, W. G. v. d. Wiel, T. Obata, and S. Tarucha, 
{\it Coherent Single Electron Spin Control in a Slanting Zeeman Field},
Phys. Rev. Lett. {\bf 96}, 047202 (2006).

\bibitem{2008LadriereNatPhys}
M. Pioro-Ladriere, T. Obata, Y. Tokura, Y.-S. Shin, T. Kubo, K. Yoshida, T. Taniyama, and S. Tarucha,
{\it Electrically driven single-electron spin resonance in a slanting Zeeman field},
Nat. Phys. {\bf 4}, 776 (2008).

\bibitem{2014YonedaPRL}
J. Yoneda, T. Otsuka, T. Nakajima, T. Takakura, T. Obata, M. Pioro-Ladriere, H. Lu, C. J. Palmstrom, A. C. Gossard, and S. Tarucha,
{\it Fast Electrical Control of Single Electron Spins in Quantum Dots with Vanishing Influence from Nuclear Spins},
Phys. Rev. Lett. {\bf 113}, 267601 (2014).

\bibitem{2005PettaSci}
J. R. Petta, A. C. Johnson, J. M. Taylor, E. A. Laird, A. Yacoby, M. D. Lukin, C. M. Marcus, M. P. Hanson, and A. C. Gossard,
{\it Coherent Manipulation of Coupled Electron Spins in Semiconductor Quantum Dots},
Science {\bf 309}, 2180 (2005).

\bibitem{2011BrunnerPRL}
R. Brunner, Y. S. Shin, T. Obata, M. Pioro-Ladriere, T. Kubo, K. Yoshida, T. Taniyama, Y. Tokura, and S. Tarucha,
{\it Two-Qubit Gate of Combined Single-Spin Rotation and Interdot Spin Exchange in a Double Quantum Dot},
Phys. Rev. Lett. {\bf 107}, 146801 (2011).

\bibitem{2013KobayashiArx}
T. Kobayashi, T. Ota, S. Sasaki, and K. Muraki,
{\it Cooperative Lifting of Spin Blockade in a Three-Terminal Triple Quantum Dot},
arXiv:1311.6582 (2013).

\bibitem{2013AmahaPRL}
S. Amaha, W. Izumida, T. Hatano, S. Teraoka, S. Tarucha, J. A. Gupta, and D. G. Austing,
{\it Two- and Three-Electron Pauli Spin Blockade in Series-Coupled Triple Quantum Dots},
Phys. Rev. Lett. {\bf 110}, 016803 (2013).

\bibitem{2010LairdPRB}
E. A. Laird, J. M. Taylor, D. P. DiVincenzo, C. M. Marcus, M. P. Hanson, and A. C. Gossard, 
{\it Coherent spin manipulation in an exchange-only qubit},
Phys. Rev. B {\bf 82}, 075403 (2010).

\bibitem{2012GaudreauNatPhys}
L. Gaudreau, G. Granger, A. Kam, G. C. Aers, S. A. Studenikin, P. Zawadzki, M. Pioro-Ladriere, Z. R. Wasilewski, and A. S. Sachrajda,
{\it Coherent control of three-spin states in a triple quantum dot},
Nat. Phys. {\bf 8}, 54 (2012).

\bibitem{2013MedfordNatNano}
J. Medford, J. Beil, J. M. Taylor, S. D. Bartlett, A. C. Doherty, E. I. Rashba, D. P. DiVincenzo, LuH, A. C. Gossard, and C. M. Marcus,
{\it Self-consistent measurement and state tomography of an exchange-only spin qubit},
Nat. Nano. {\bf 8}, 654 (2013).

\bibitem{2010TakakuraAPL}
T. Takakura, M. Pioro-Ladriere, T. Obata, Y. S. Shin, R. Brunner, K. Yoshida, T. Taniyama, and S. Tarucha,
{\it Triple quantum dot device designed for three spin qubits},
Appl. Phys. Lett. {\bf 97}, 212104 (2010).

\bibitem{2015Nakajima}
T. Nakajima, M. R. Delbecq, T. Otsuka, S. Amaha, J. Yoneda, A. Noiri, K. Takeda, G. Allison, A. Ludwig, A. D. Wieck, and Seigo Tarucha,
in preparation.

\bibitem{2012ThalineauAPL}
R. Thalineau, S. Hermelin, A. D. Wieck, C. Bauerle, L. Saminadayar, and T. Meunier,
{\it A few-electron quadruple quantum dot in a closed loop},
Appl. Phys. Lett. {\bf 101}, 103102 (2012).

\bibitem{2014TakakuraAPL}
T. Takakura, A. Noiri, T. Obata, T. Otsuka, J. Yoneda, K. Yoshida, and S. Tarucha, 
{\it Single to quadruple quantum dots with tunable tunnel couplings},
Appl. Phys. Lett. {\bf 104}, 113109 (2014).

\bibitem{2012ShulmanSci}
M. D. Shulman, O. E. Dial, S. P. Harvey, H. Bluhm, V. Umansky, and A. Yacoby, 
{\it Demonstration of Entanglement of Electrostatically Coupled Singlet-Triplet Qubits},
Science {\bf 336}, 202 (2014).

\bibitem{2014DelbecqAPL}
M. R. Delbecq, T. Nakajima, T. Otsuka, S. Amaha, J. D. Watson, M. J. Manfra, and S. Tarucha,
{\it Full control of quadruple quantum dot circuit charge states in the single electron regime},
Appl. Phys. Lett. {\bf 104}, 183111 (2014).

\bibitem{2010BarthelPRB}
C. Barthel, M. Kjargaard, J. Medford, M. Stopa, C. M. Marcus, M. P. Hanson, and A. C. Gossard,
{\it Fast sensing of double-dot charge arrangement and spin state with a radio-frequency sensor quantum dot},
Phys. Rev. B {\bf 81}, 161308 (2010).

\bibitem{1998SchelkopfSci}
R. J. Schoelkopf, P. Wahlgren, A. A. Kozhevnikov, P. Delsing, and D. E. Prober,
{\it The Radio-Frequency Single-Electron Transistor (RF-SET): A Fast and Ultrasensitive Electrometer},
Science {\bf 280}, 1238 (1998).

\bibitem{2007ReillyAPL}
D. J. Reilly, C. M. Marcus, M. P. Hanson, and A. C. Gossard,
{\it Fast single-charge sensing with a rf quantum point contact},
Appl. Phys. Lett. {\bf 91}, 162101 (2007).

\bibitem{2005KoppensSci}
F. H. L. Koppens, J. A. Folk, J. M. Elzerman, R. Hanson, L. H. W. van Beveren, I. T. Vink, H. P. Tranitz, W. Wegscheider, L. P. Kouwenhoven, and L. M. K. Vandersypen,
{\it Control and Detection of Singlet-Triplet Mixing in a Random Nuclear Field},
Science {\bf 309}, 1346 (2005).

\bibitem{2014ChesiPRB}
S. Chesi, Y.-D. Wang, J. Yoneda, T. Otsuka, S. Tarucha, and D. Loss,
{\it Single-spin manipulation in a double quantum dot in the field of a micromagnet},
Phys. Rev. B {\bf 90}, 235311 (2014).

\bibitem{2009BarthelPRL}
C. Barthel, D. J. Reilly, C. M. Marcus, M. P. Hanson, and A. C. Gossard,
{\it Rapid Single-Shot Measurement of a Singlet-Triplet Qubit},
Phys. Rev. Lett. {\bf 103}, 160503 (2009).

\bibitem{2003PotokPRL}
R. M. Potok, J. A. Folk, C. M. Marcus, V. Umansky, M. Hanson, and A. C. Gossard,
{\it Spin and Polarized Current from Coulomb Blockaded Quantum Dots},
Phys. Rev. Lett. {\bf 91}, 016802 (2003).

\bibitem{2003HansonPRL}
R. Hanson, B. Witkamp, L. M. K. Vandersypen, L. H. W. van Beveren, J. M. Elzerman, and L. P. Kouwenhoven,
{\it Zeeman Energy and Spin Relaxation in a One-Electron Quantum Dot},
Phys. Rev. Lett. {\bf 91}, 196802 (2003).

\bibitem{2005BeverenNJP}
L. H. W. v. Beveren, R. Hanson, I. T. Vink, F. H. L. Koppens, L. P. Kouwenhoven, and L. M. K. Vandersypen,
{\it Spin filling of a quantum dot derived from excited-state spectroscopy},
New J. Phys. {\bf 7}, 182 (2005).

\bibitem{2015YonedaAPE}
J. Yoneda, T. Otsuka, T. Takakura, M. Pioro-Ladriere, R. Brunner, H. Lu, T. Nakajima, T. Obata, A. Noiri,  C. J. Palmstrom, A. C. Gossard, and S. Tarucha,
{\it Robust micromagnet design for fast electrical manipulations of single spins in quantum dots},
Appl. Phys. Exp. {\bf 8}, 084401 (2015).

\end{references}
\end{document}